\documentclass[epj]{webofc}
\usepackage[utf8]{inputenc}
\usepackage[varg]{txfonts}   
\usepackage{booktabs,ulem}
\usepackage{xcolor}
\definecolor{darkred}{rgb}{0.4,0.0,0.0}
\definecolor{darkgreen}{rgb}{0.0,0.4,0.0}
\definecolor{darkblue}{rgb}{0.0,0.0,0.4}
\usepackage[bookmarks,linktocpage,colorlinks,
    linkcolor = darkred,
    urlcolor  = darkblue,
    citecolor = darkgreen]{hyperref}

\wocname{EPJ Web of Conferences}
\woctitle{Lattice2017}
\begin{document}
%
\selectlanguage{english}
\title{%
Investigating BSM Models with Large Scale Separation
}
\author{%
\firstname{Anna} \lastname{Hasenfratz}\inst{1} \and
\firstname{Claudio} \lastname{Rebbi}\inst{2}\fnsep\thanks{Speaker, 
\email{rebbi@bu.edu}} \and
\firstname{Oliver}  \lastname{Witzel}\inst{3}\thanks{Present address: Department of Physics, University of Colorado, Boulder, CO 80309, USA}
}
\institute{%
Department of Physics, University of Colorado, Boulder, CO 80309, USA
\and
Department of Physics and Center for Computational Science, Boston University, Boston, MA 02215, USA
\and
Higgs Centre for Theoretical Physics, University of Edinburgh, Edinburgh, EH9 3FD, UK
}
\abstract{%
Mass-split systems based on a conformal infrared fixed point provide
a low-energy effective description of beyond the standard model  
systems with large scale
separation. We report results of exploratory investigations with
four light and eight heavy flavors using staggered fermions, and up
to five different values for the light flavor mass, five different
heavy flavor masses, and two values of the bare gauge coupling.
}
\maketitle
\section{Introduction}\label{intro}

We focus on ``beyond the standard model'' (BSM) theories based on 
strong dynamics. These theories assume the existence of a strongly 
coupled system of gauge fields and fermions with a broken chiral 
symmetry: three Goldstone bosons provide the longitudinal components
of the $W$'s and $Z$, while the Higgs emerges either
as a pseudo-Goldstone boson or as a dilaton. Experiment demands
that a viable theory accommodates a large separation between
the scale of electroweak symmetry breaking and the scale of
excitations in the new strongly coupled sector.  This is achieved
by formulating the strongly coupled theory in such a way that
it lies within, or close to, the conformal window, i.e.~the range of parameters
for which a theory has an infrared fixed point (IRFP) but is still 
asymptotically free (see e.g.~\cite{Luty:2004ye,Dietrich:2006cm,Vecchi:2015fma,Ferretti:2013kya,Ma:2015gra}).  Now, while with some
gauge groups a theory can be inside the conformal window
with few fermions, in other cases, and most notably
with the SU(3) gauge group, a large number of fermions
is required to get a model with an IRFP.
In this latter case only a small subset of the fermions
can have vanishing mass:  the other fermions must be given
a mass to avoid the occurrence of way too many Goldstone
bosons.  Thus the theory becomes a ``mass broken conformal
theory'', that is a theory which would have an IRFP with all
fermions massless, but which is kept away from the IRFP 
by giving mass to a subset of its fermions.  Purpose of this
paper is to study the properties of such ``mass broken
conformal theories''.  We will proceed by presenting first
some unproven, but plausible theoretical considerations based on non-perturbative 
Wilsonian renormalization group arguments,
following them with the illustration of numerical results
which appear to validate our  assumptions.  The conclusion
will be that ``mass broken conformal theories'' have
remarkable properties of universality, not unlike QCD
where with massless quarks the theory is independent
of any coupling constant.

\section{Theoretical Considerations}\label{theory}

In this section we will formulate some  theoretical expectations.
We consider an SU(3) lattice gauge theory with $N_f$ identical fermions.
We take the lattice to be of infinite extent, we denote by $\beta$,
in the standard manner, a parameter proportional to the inverse
gauge coupling squared, denote by $am$ the
mass of the fermions in lattice units\footnote{In this paper we follow
the convention of using symbols such as $m$, $M$, $F_\pi$ to denote 
dimensionful quantities. Correspondingly the  mass of the fermions in 
lattice units must be denoted by $am$, $a$ being the lattice spacing,
although in absence of scale setting, neither $a$ nor $m$ are
separately defined.}, and assume that the fermions
have been discretized in a manner which preserves chiral symmetry for
$am=0$ and avoids  unwanted fermionic copies. 
When we mention an observable,
we take this to be the actual value taken by the observable, which
is in principle theoretically well defined.

For our discussion, let us take first $N_f=4$.  The fermion mass
$am$ can be taken to zero.  The theory exhibits chiral symmetry
breaking.  Observables such as the pseudoscalar decay constant $aF_\pi$,
masses $aM$ etc.~have well defined finite values in lattice units which depend 
only on $\beta$.  The continuum limit is recovered by letting 
$\beta \to \infty$: in this limit dimensionless ratios of observables
approach their continuum values, any non-vanishing observable 
can be used to set the scale and all other observables are
fully determined in a manner independent of any free parameter
(dimensional transmutation.) We may take one or two 
masses, $am_1, am_2,$ different from zero.  
As $\beta \to \infty$ to recover the continuum limit
we  must also send $am_1, am_2 \to 0$ in a controlled
manner.  The spectrum of masses and other observables will depend 
now on the way $am_1, am_2 $ go to zero as 
$\beta \to \infty$ and the ratios of observables
to $F_\pi $ will tend to finite limits.  
We recover QCD with massless up and down quarks and 
non-vanishing strange and charm masses. 
Nothing new here:  we described  a QCD-like theory with  massless up and down 
quarks and non-vanishing strange and charm quark masses.

Let us consider then the SU(3) theory with 12 fermions of mass
$am=0$.  We assume that the theory is conformal. The correlations 
functions will exhibit non-trivial power law behavior at 
large distances.  If we rescale lengths 
appropriately, the long range behavior with any two values of 
$\beta$ will be the same, because under renormalization
group transformations the theory runs to to an IRFP, in the space of 
infinite couplings. (The location of the IRFP will depend on the 
specifics of the renormalization.)  The continuum behavior of the
correlation functions is recovered by going to very large
lattice distances.  If we are close enough to the IRFP, 
the lattice will approximate well the continuum also at moderate
and small distances: then the short distance behavior of the
correlation functions will be trivial because 
the theory is asymptotically free in the UV.

Assume now that we keep 4 fermions massless and give a mass $am$ 
to the other 8 fermions \cite{Brower:2014dfa,Brower:2015owo,Hasenfratz:2016gut}. The observables (in lattice units) will now 
depend on $\beta$ and $am$.  As in the case with $N_f=4$,
at large distances the correlation functions will exhibit
exponential behavior and we can measure a non-vanishing 
$aF_\pi$ as well as non-vanishing masses 
and other observables.  
In a renormalization transformation, 
in the space of infinite couplings, the system will move away from the 
IRFP with $am=0$ because $am$ 
is a relevant parameter.   To recover the continuum limit with 
$aF_\pi \to 0$ we must let $am$ approach its fixed-point value $am=0$, 
but, and here is the crucial point, contrary to 
the case of the theory with $N_f=4$,
now we do not need to let $\beta \to \infty$, since, 
as $am \to 0$, the theory will flow closer and
closer to the IRFP no matter what  the original 
$\beta$ is.
If we take $am$ sufficiently
small, the ratios of observables, whether built out of the massless
fermions, massive fermions or both, and in particular the ratios of
observables to $F_\pi$, will be independent of 
$am$ and  $\beta$.  
If we give two different masses, $am_1, am_2$, to 
the massive fermions, or if we give a small mass 
$am_\ell$ to the massless fermions, 
hyperscaling arguments~\cite{Hasenfratz:2016uar,Hasenfratz:2016gut}
show that the ratios of observables to
$F_\pi$ or among themselves only depend 
on the ratios of the bare Lagrangian masses.
  
In conclusion, we can take $F_\pi$ 
to set the scale, and then the continuum theory is fully defined 
by the ratio of Lagrangian masses.
In particular, if we only have a common mass parameter $am$
for the 8 massive fermions, while the 4 light fermions 
are kept at zero mass, then the theory is fully defined without any 
free parameter.  In this case $am$ plays a role 
similar to $\beta$ for massless QCD.

\section{Numerical Results}\label{numerical}

In this section we illustrate numerical results which appear
to validate the theoretical arguments presented above.
We simulated a theory \cite{Brower:2014dfa,Brower:2015owo,Hasenfratz:2016gut} with:
\begin{itemize} 
\item one staggered field (= 4 flavors) with light mass 
${m_\ell}$ plus two staggered fields (= 8 flavors) with
heavy mass ${m_h}$.   The simulations have been done done with
${am_\ell=0.003, 0.005, 0.010, 0.015 ,0.025, 0.035}$,
${am_h=0.050, 0.060, 0.080, 0.100}$; 
\item a fundamental-adjoint gauge action with ${\beta=4.0, 
\beta_a=-\beta/4}$  \cite{Cheng:2013bca,Cheng:2013xha},
and {nHYP} smeared staggered fermions \cite{Hasenfratz:2001hp,Hasenfratz:2007rf} ;
\item lattice sizes mostly ${24^3 \times 48}$ and 
${32^3 \times 64}$, but also 
${16^3 \times 32}$~(exploratory), 
${36^3 \times 64}$ and ${48^3 \times 96}$.
\item We also simulated a
system with ${\beta=4.4}$, 
${a m_h=0.070}$, 
${a m_\ell=0.009, 0,013125, 0.0175, 0.0245}$, on 
a ${32^3 \times 64}$ lattice.
\end{itemize}
\begin{figure}[h!] 
  \centering
\includegraphics[width=0.6\textwidth]{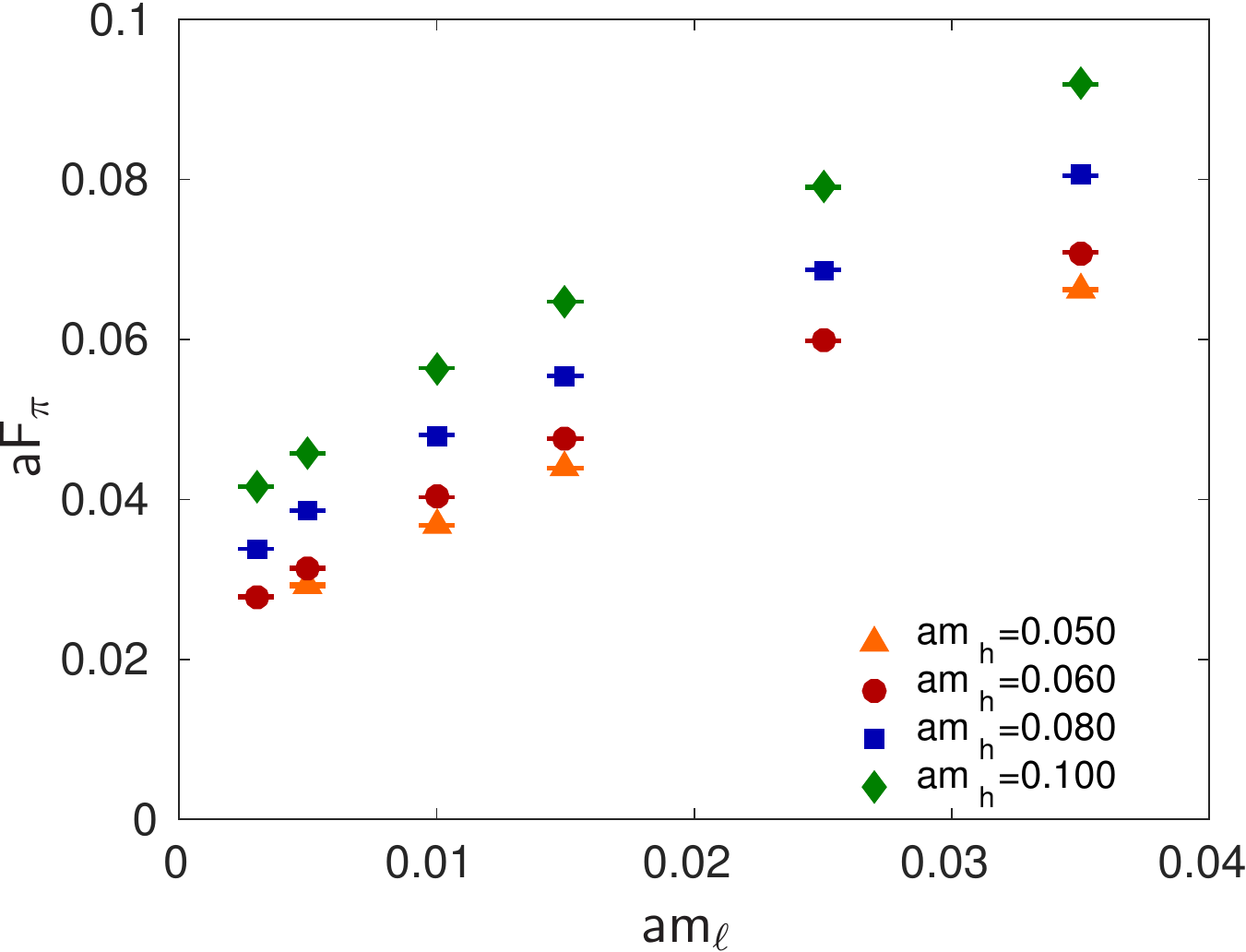}  
\caption{The pseudoscalar decay constant in lattice units. }
  \label{fig1}
\end{figure}

In Figure~\ref{fig1} we present  the value of the pseudoscalar decay
constant, $aF_\pi$ in lattice units. $aF_\pi$ is seen to decrease
as $a m_h$ decreases, following the expectation that it should
go to zero as $a m_h$ approaches the fixed-point value $a m_h=0$. 
\begin{figure}[ht] 
  \centering
\includegraphics[height=0.25\textheight]{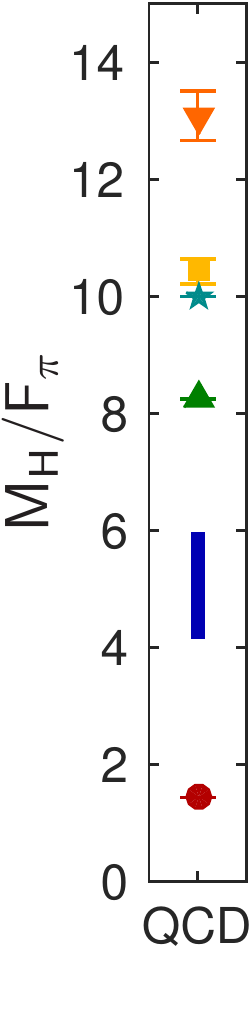}
\includegraphics[height=0.25\textheight]{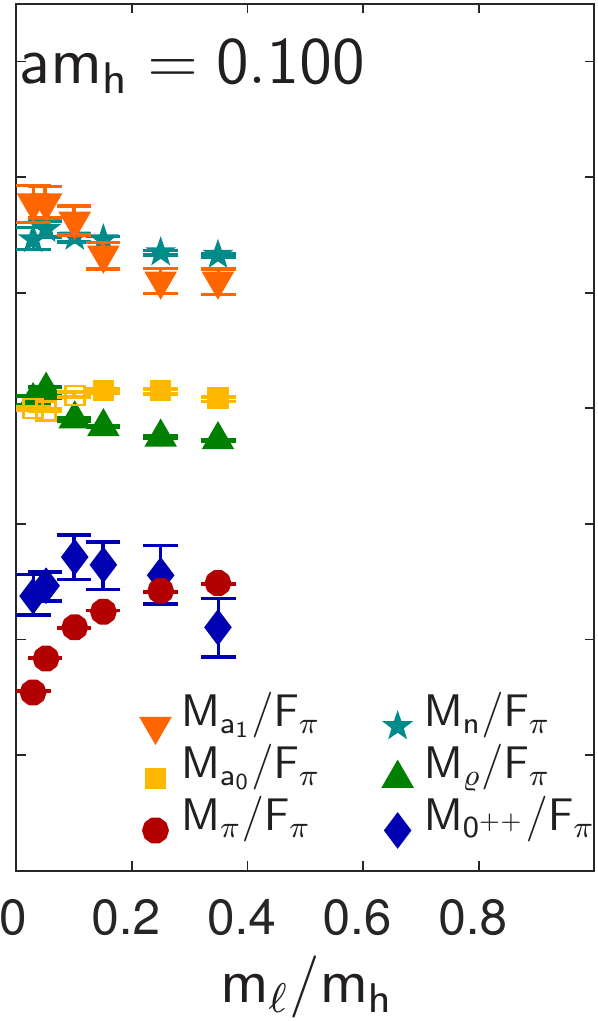}
\includegraphics[height=0.25\textheight]{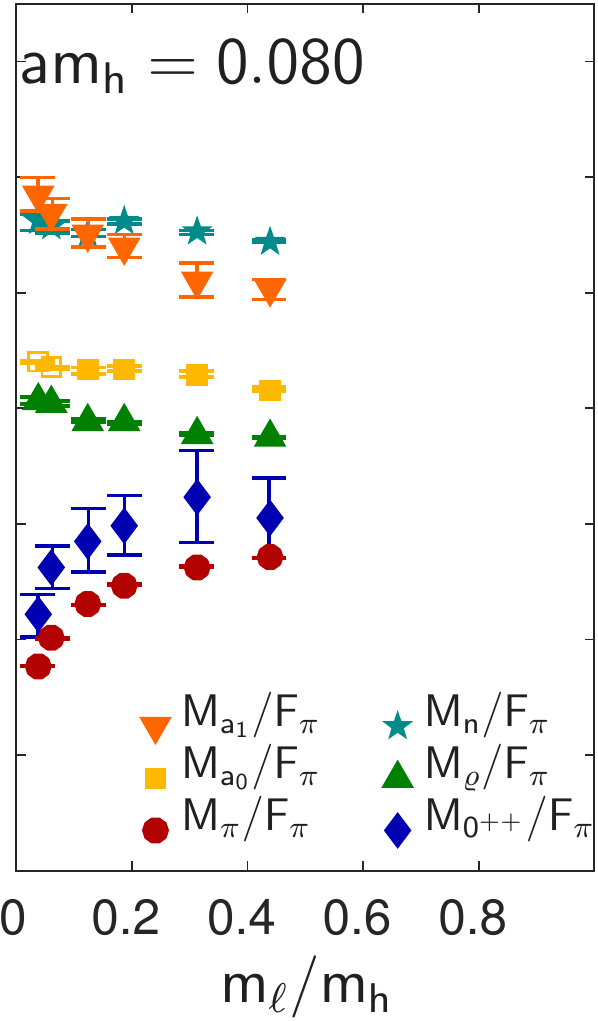}
\includegraphics[height=0.25\textheight]{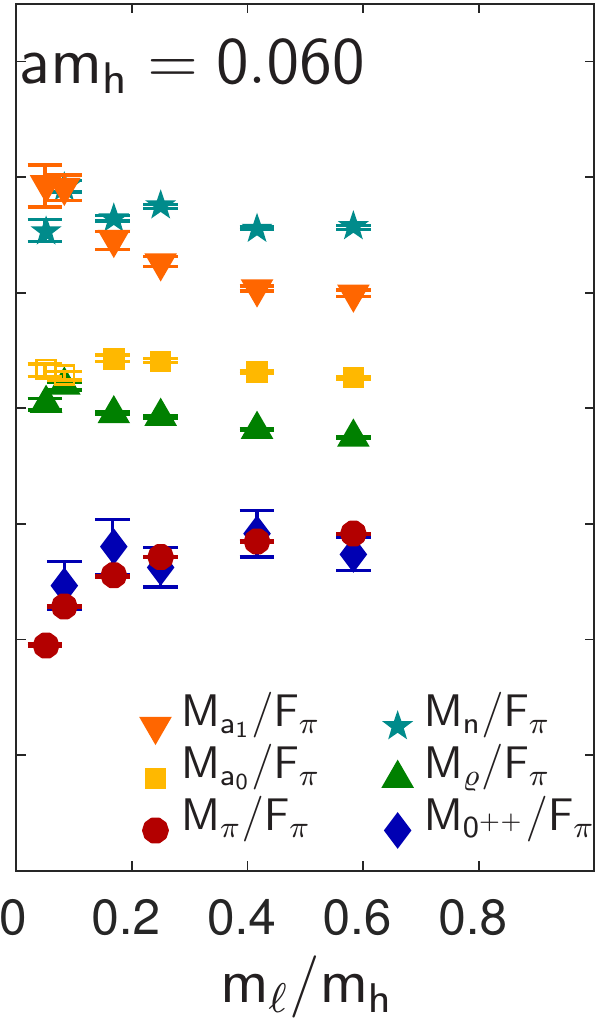}
\includegraphics[height=0.25\textheight]{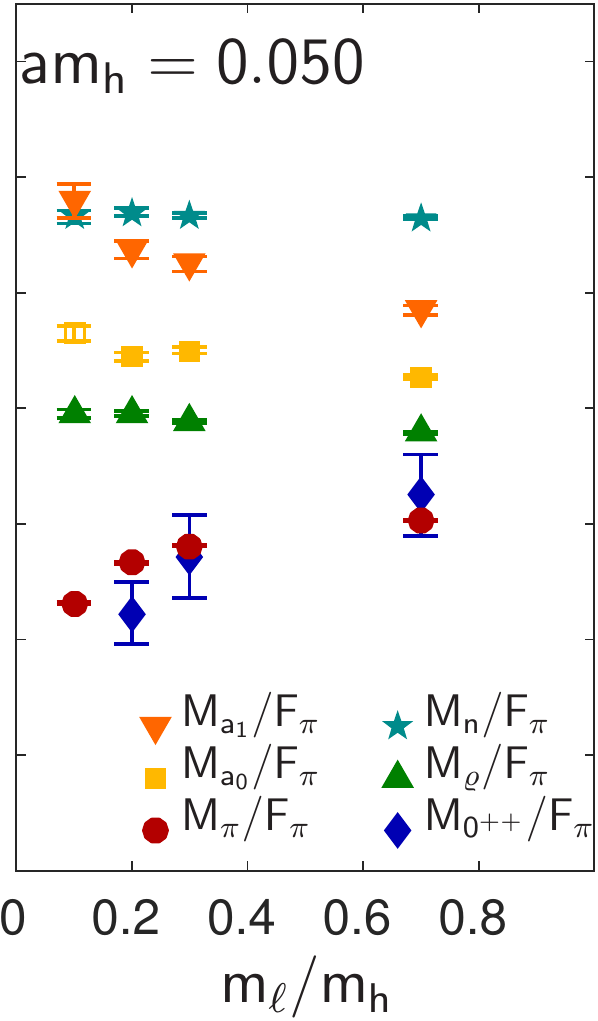}
\includegraphics[height=0.25\textheight]{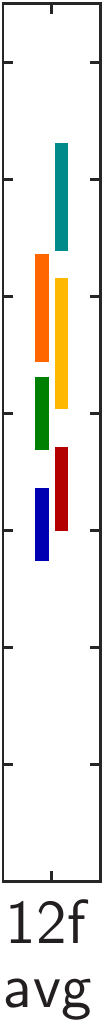}
\caption{Ratios of the masses of $\pi,\, \rho,\, a_0,\, a_1,\, n$, and of 
the $0^{++}$ over $F_\pi$.
The left panel shows results for QCD taken from the PDG \cite{Agashe:2014kda}, the right panel averages for a (mass deformed) 12 flavor theory \cite{Aoki:2012eq,Fodor:2011tu,Cheng:2013xha,Aoki:2013zsa}.}
  \label{fig2}
\end{figure}

 Our light spectrum results for $\beta=4.0$ are summarized in the four panels 
of Fig~\ref{fig2}. (In this figure, as well as in the following
figures, we use the symbols $\pi,\, \rho,\, \dots$ as a short-hand
for pseudoscalar, vector etc., referring to the particle 
state with the same quantum numbers as in QCD.) We plot ratios of 
hadron masses $M_H$ over $F_\pi$, and as functions
of $m_\ell/m_h$, in order to remove the scale dependence.  
We note that $M_\pi/F_\pi$ decreases 
towards zero for $m_\ell/m_h \to 0$, giving evidence of chiral symmetry 
breaking.  We also observe that the $0^{++}$ is light, with
$M_{0^{++}}< M_\rho$, tracking the pion.

\begin{figure}[h!] 
  \centering
\includegraphics[height=0.3\textheight]{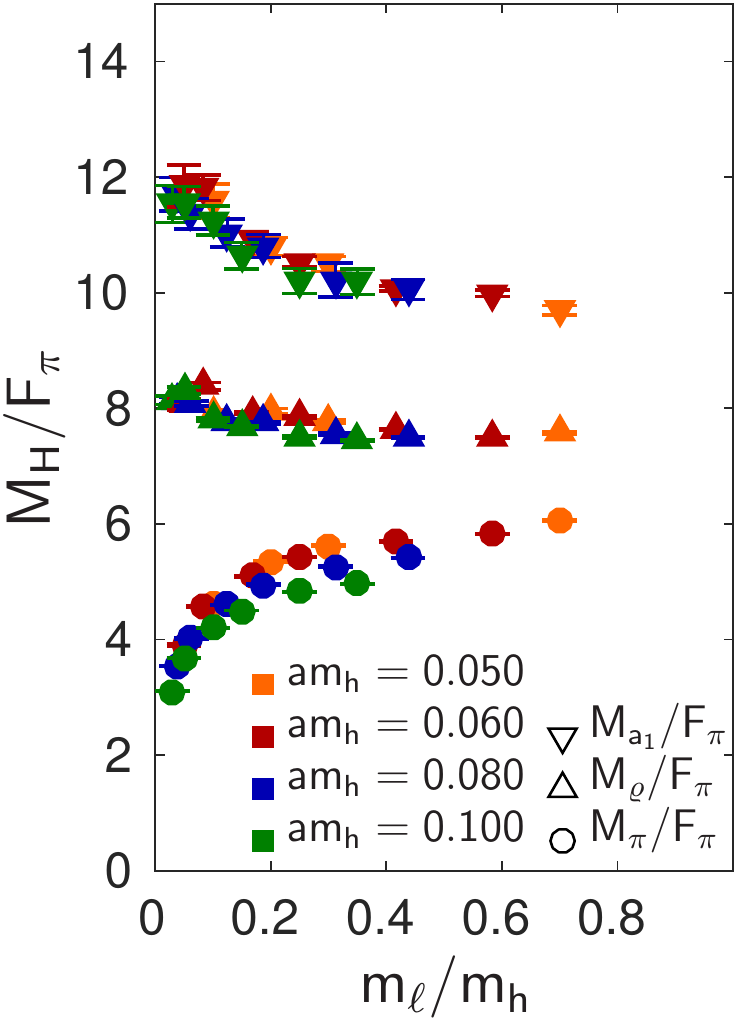}
\hskip 14mm
\includegraphics[height=0.3\textheight]{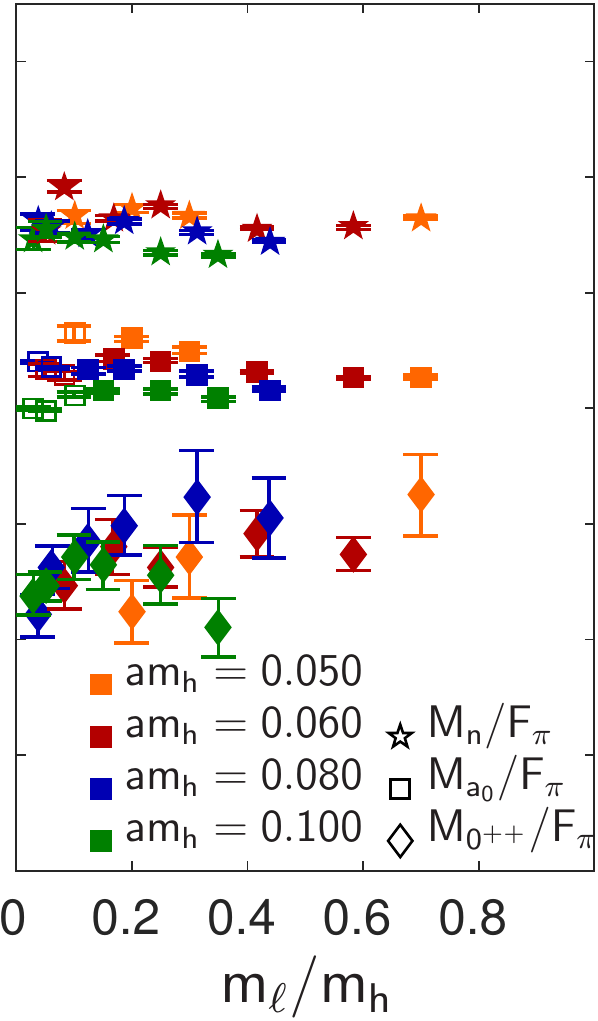}
\caption{Hyperscaling: light hadron masses. Data for different
$m_h$ are superimposed.}
  \label{fig3}
\end{figure}

In Figure~\ref{fig3} we superimpose the data obtained with different
values of $m_h$.  Hyperscaling arguments \cite{Hasenfratz:2016uar,Hasenfratz:2016gut} 
indicate that in the vicinity of the IRFP all scale independent
quantities will be only a function of $m_\ell/m_h$, with no separate
dependence on $m_h$.  The way in which the data appear to line-up 
in Fig.~\ref{fig3} provides strong evidence for hyperscaling.

\begin{figure}[ht!] 
  \centering
\includegraphics[height=0.3\textheight]{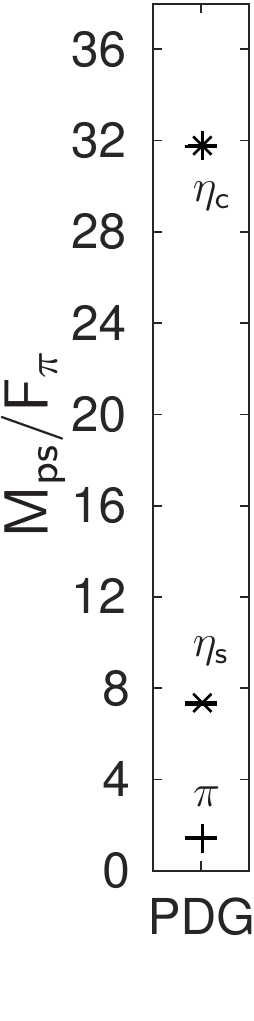}
\includegraphics[height=0.3\textheight]{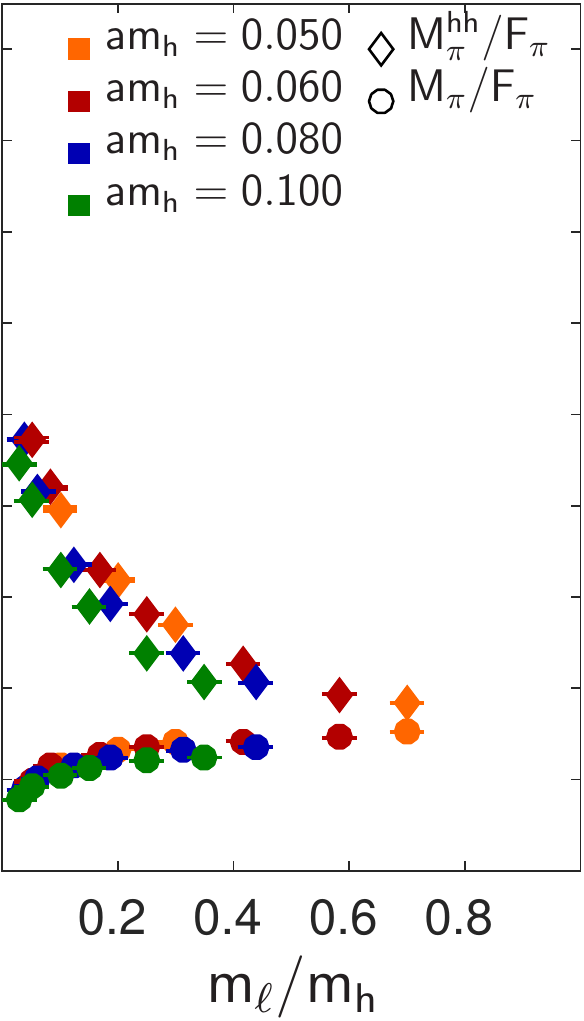}
\includegraphics[height=0.3\textheight]{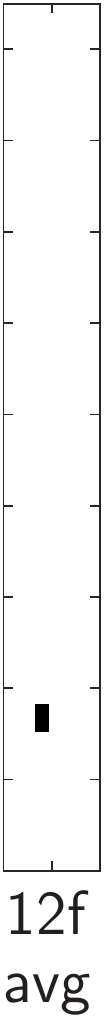}
\includegraphics[height=0.3\textheight]{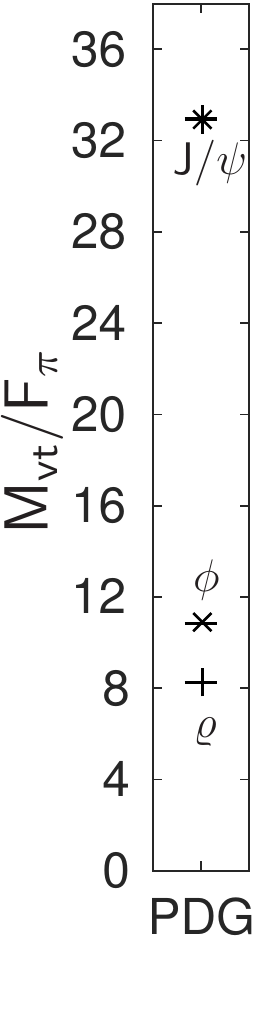}
\includegraphics[height=0.3\textheight]{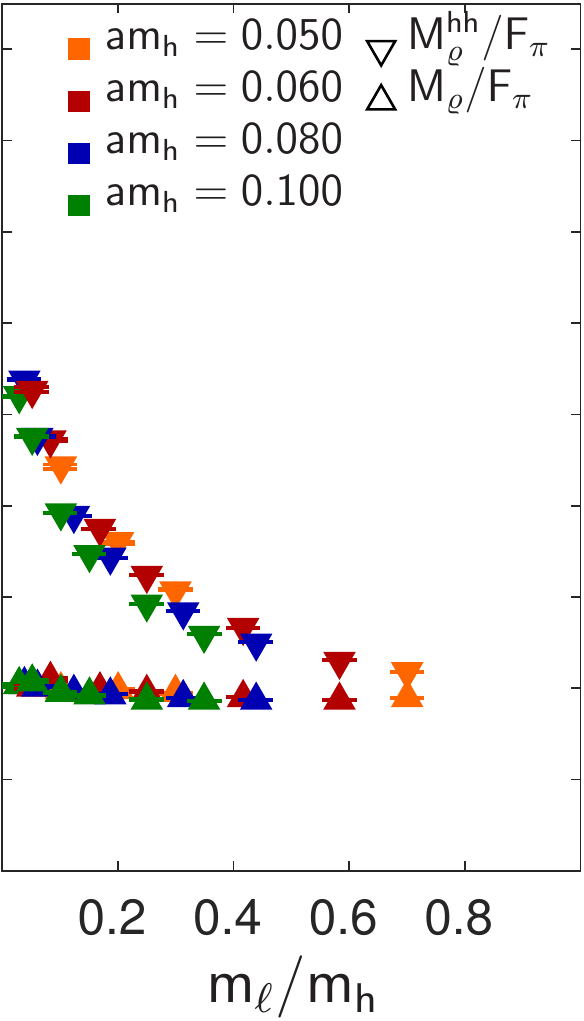}
\includegraphics[height=0.3\textheight]{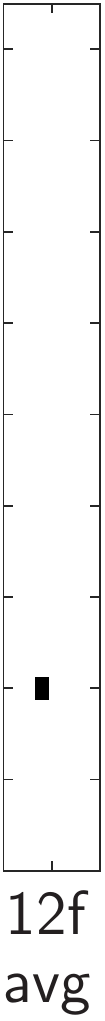}
\vskip 0.2cm
\includegraphics[height=0.3\textheight]{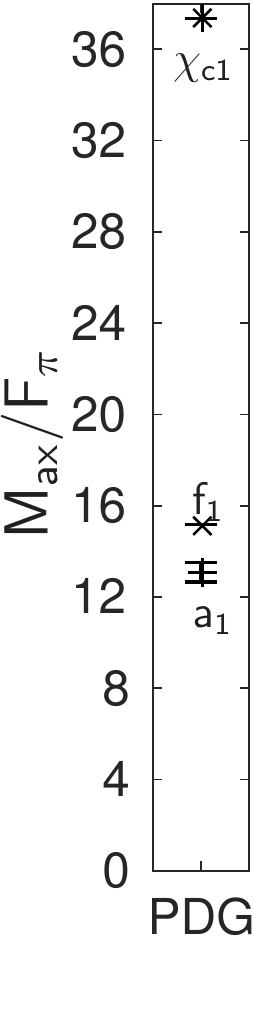}
\includegraphics[height=0.3\textheight]{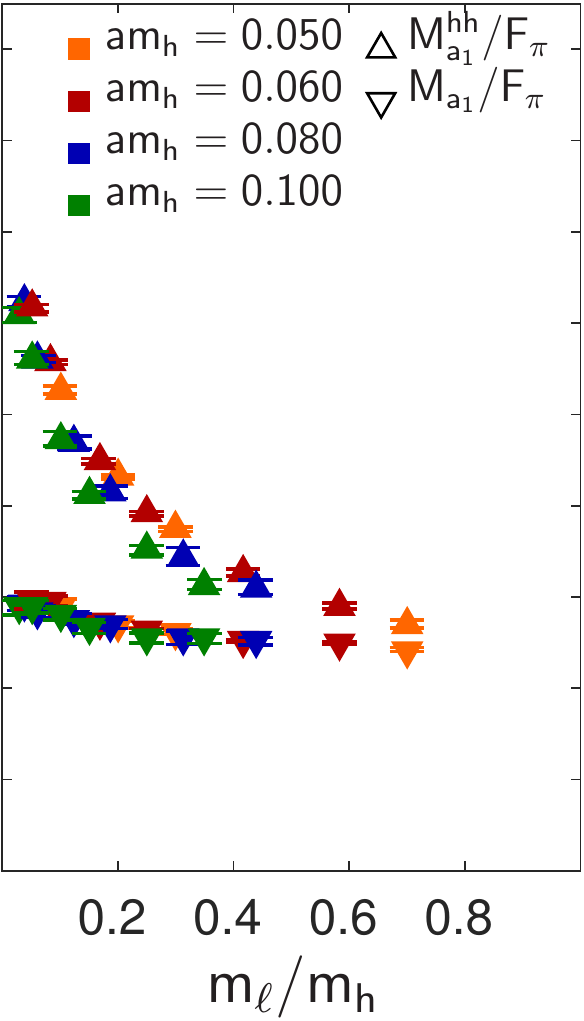}
\includegraphics[height=0.3\textheight]{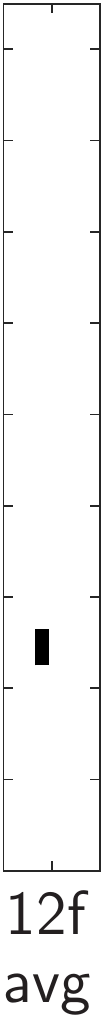}
\caption{Ratios of the masses of $\pi,\, \rho,\, a_1$ over $F_\pi$ for
light-light and heavy-heavy composites.
The left panels shows result for QCD \cite{Agashe:2014kda} with the (unphysical) $\eta_s$ taken from \cite{Dowdall:2013rya}, the right panels for a (mass deformed)
12 flavor theory \cite{Aoki:2012eq,Fodor:2011tu,Cheng:2013xha,Aoki:2013zsa}.}
  \label{fig4}
\end{figure}

We also calculated the spectrum of masses  for states made of 
heavy fermions.  The spectra of light-light and heavy-heavy
meson composites is illustrated in the the three panels of Fig.~\ref{fig4}.
The heavy-heavy spectrum also gives clear evidence of hyperscaling.

The upward bend in the heavy-heavy spectrum is in part due to 
the fact that  light-light $F_\pi $ shows a marked decrease for $m_\ell/m_h \to 0$.

\begin{figure}[h] 
  \centering
\includegraphics[height=0.3\textheight]{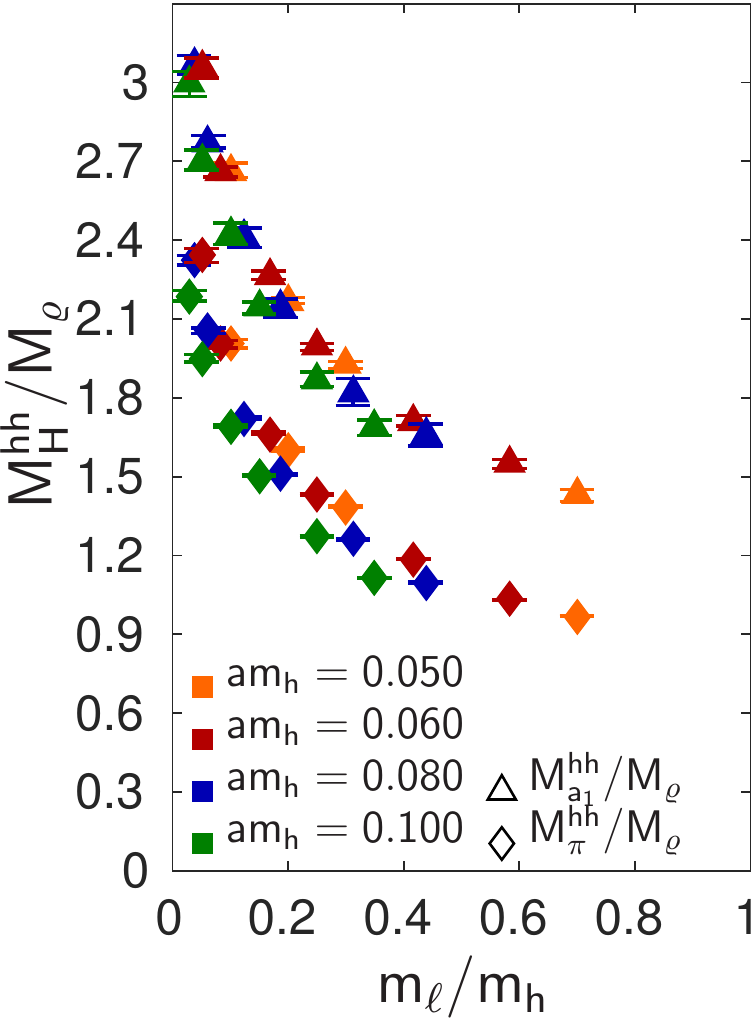}
\hskip 10mm
\includegraphics[height=0.3\textheight]{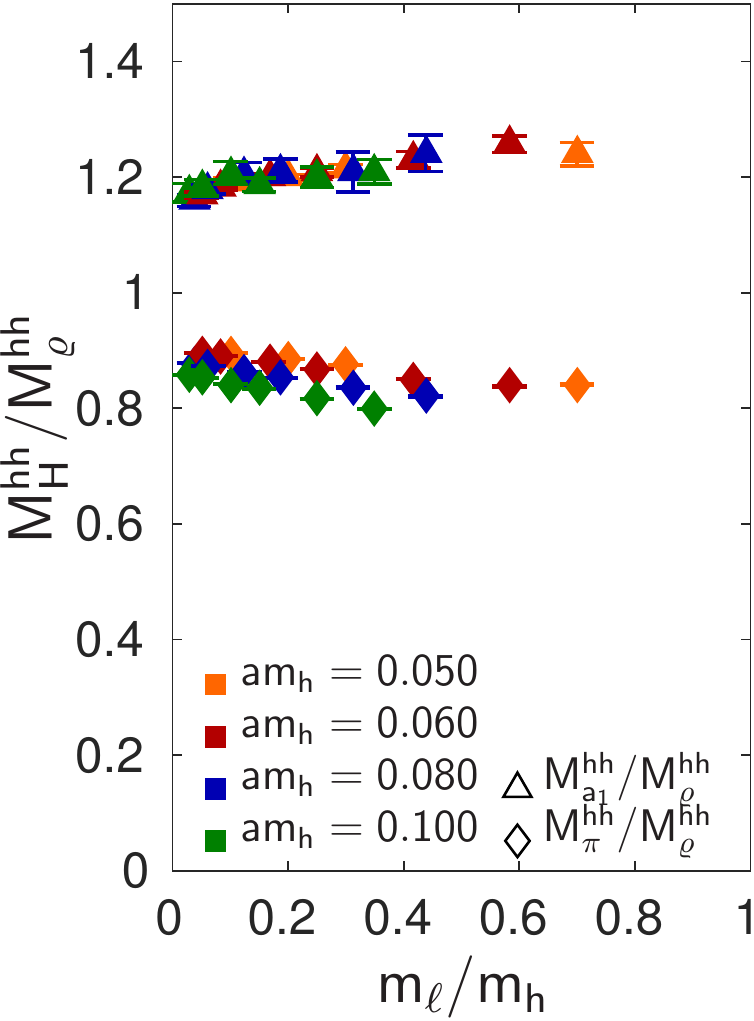}
\caption{Ratios of the heavy-heavy $\pi$ and $a_1$ masses over
the mass of the light-light  $\rho$ (left panel) and
of the heavy-heavy  $\rho$.}
  \label{fig5}
\end{figure}

In Figure~\ref{fig5} we show the ratio of the heavy-heavy $\pi$ and
$a_1$ masses over the mass of light-light and heavy-heavy $\rho$.
Once again the upward bend in the left panel can be attributed to the 
decrease of the light-light $M_\rho$ in the denominator. Ratios
against heavy-heavy $M_\rho $ show little variation.

\begin{figure}[h!] 
  \centering
\includegraphics[height=0.25\textheight]{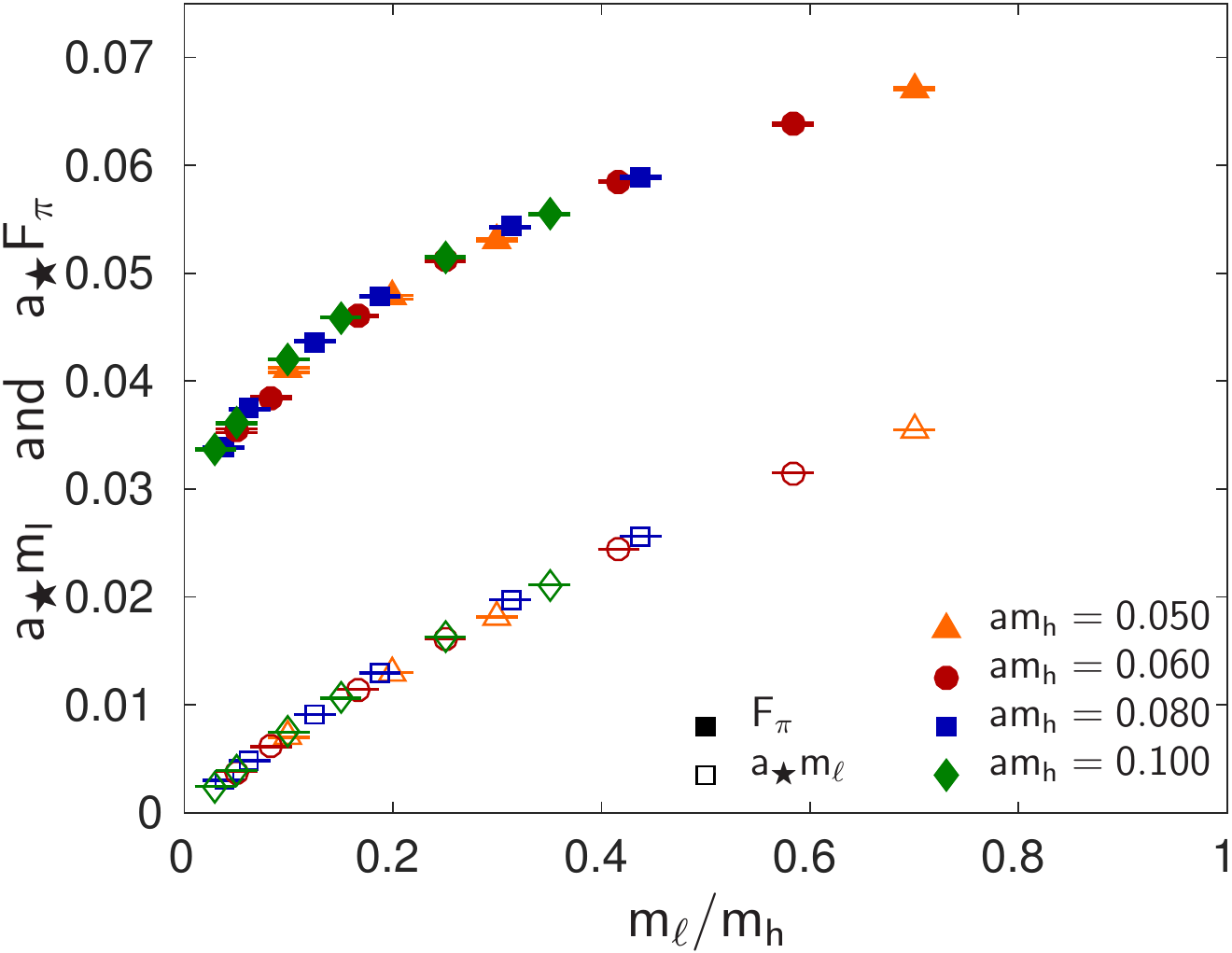}
\hskip 5mm
\includegraphics[height=0.25\textheight]{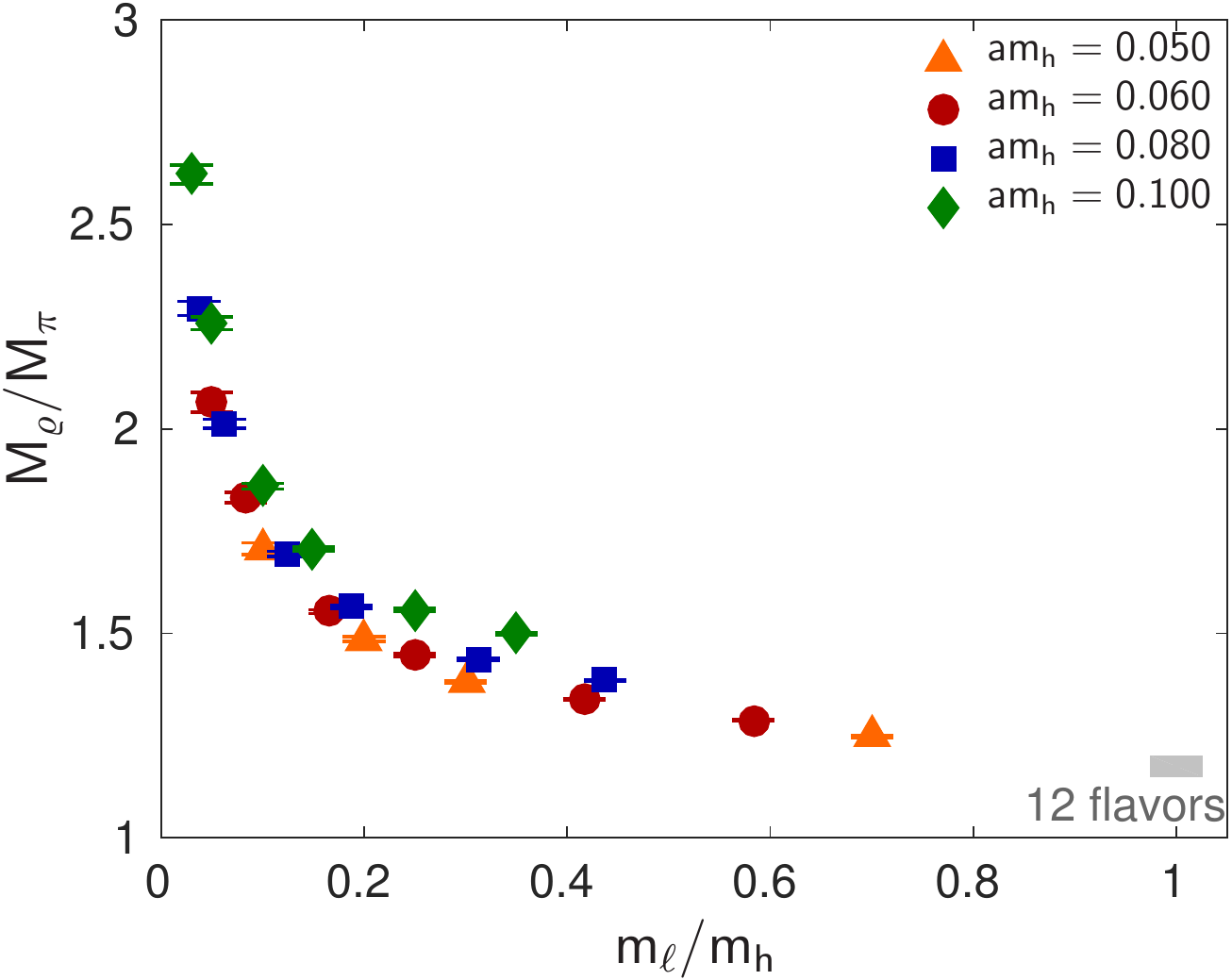}
\caption{Right panel: $F_\pi$ and the light fermion mass $m_\ell$ as 
functions of the ratio $m_\ell/m_\pi$. For this graph the scale is based
on the Wilson flow.  Left panel: Ratio of the $\rho$ mass over the $\pi$
mass.}
  \label{fig6}
\end{figure}

Figure~\ref{fig6} presents more detailed evidence for the fact that
 chiral symmetry is broken.  The left panel shows that as
the ratio $m_\ell/m_h$ goes to zero, $M_\pi$ tends to zero
while $F_\pi$ tends to a finite limit, a clear indication of chiral
symmetry breaking.  In order to display physical values for $F_\pi$,
as opposed to $aF_\pi$ (in lattice units), we multiply $F_\pi$ by
a scale, denoted by $a_\bigstar$, which represents the lattice spacing
for $\beta=4.0, am_\ell=0.003,am_h=0.080$ as determined through the
Wilson flow.  The values in lattice units, $aM_\pi, aF_\pi$
at all other values of $am_\ell, am_h$ are converted to the
values in Fig.~\ref{fig6}  multiplying them by $a_\bigstar/a$ as
determined by the Wilson flow.   The sharp increase of $M_\rho/M_\pi$
for $m_\ell/m_h\to 0$, which is seen in the right panel of Fig.~\ref{fig6},
is more evidence of chiral symmetry breaking.

\begin{figure}[h] 
  \centering
\includegraphics[height=0.3\textheight]{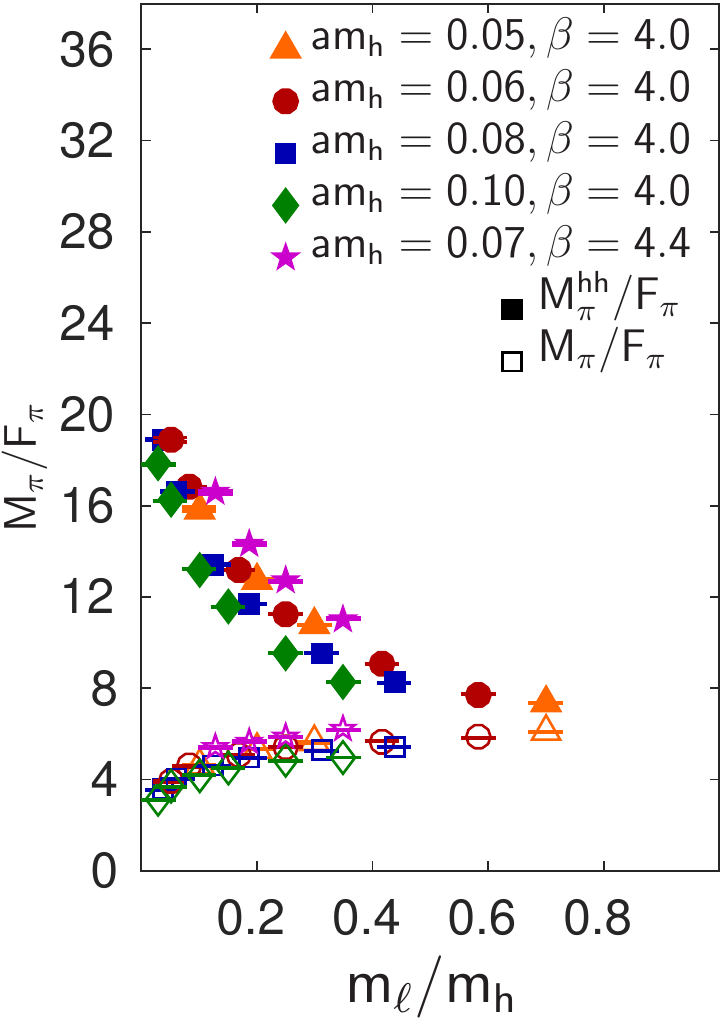}
\includegraphics[height=0.3\textheight]{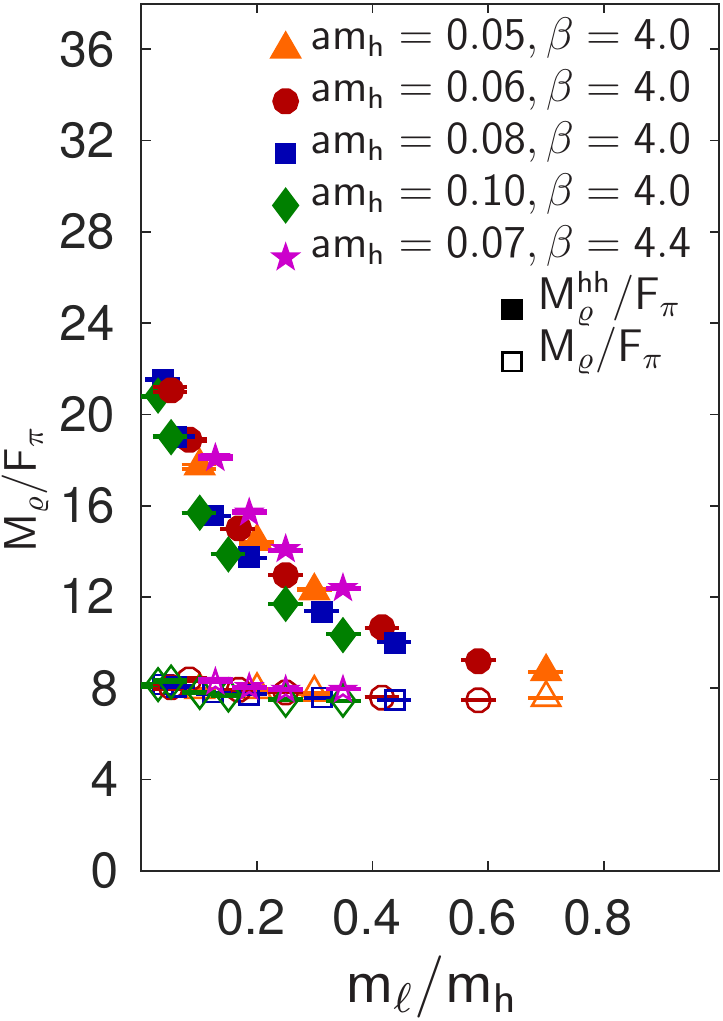}
\includegraphics[height=0.3\textheight]{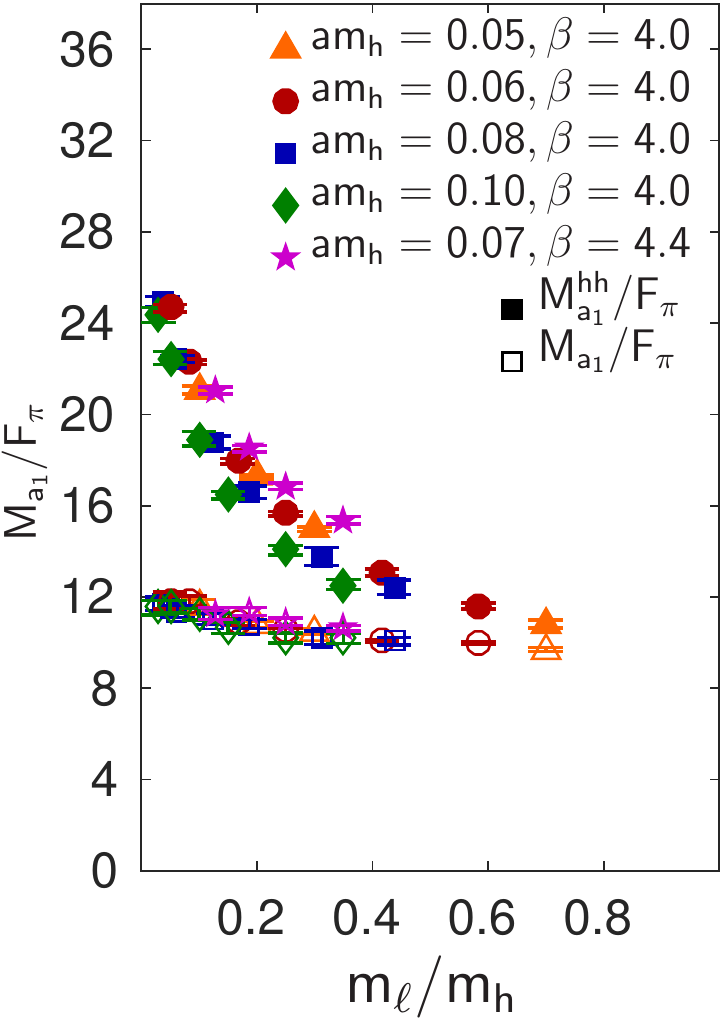}
\caption{The coupling constant $\beta$ is an irrelevant parameter:
data obtained for the light-light and heavy-heavy masses of $\pi$,
$\rho$ and $a_1$ with $\beta = 4.0, am_h=0.100,\,0.080,\,0.060,\,0.050$
and $\beta = 4.4, am_h=0.070$ are superimposed.}
  \label{fig7}
\end{figure}

In Section~\ref{theory} we stated that in the neighborhood 
of the IRFP the coupling constant $\beta$ is an irrelevant parameter.  
In order to verify that this is indeed the case  we performed additional 
simulations with $\beta = 4.4,\; am_h=0.070$.  If $\beta$ is an irrelevant
parameter, then the ratios of physical quantities should be largely 
independent of its value.  In Figure~\ref{fig7} we plot the results
for $M_\pi,\, M_\rho,\, M_{a_1}$, in units of $F_\pi$, obtained from all
our simulations.  We already observed how the data corresponding
to different $m_h$ at $\beta=4.0$ fall on top of each other,
confirming hyperscaling.  From Figure~\ref{fig7} we see that also
the data obtained with $\beta=4.4$ follow the same pattern. 
While the small deviations from perfect alignment  
can be explained by scaling violations, finite size effects, 
and other systematic factors, the way in which all the data 
appear to line up, showing a common dependence on $m_\ell/m_h$, is
impressive and appear to validate the theoretical expectations
expounded in Sect.~\ref{theory}.

\section{Conclusions and Outlook}\label{outlook}

Our results give evidence for hyperscaling in the
heavy fermion masses and for the irrelevance of the gauge
coupling:  for sufficiently small $m_h$,
ratios of physical quantities will depend only on
$m_\ell/m_h$ and will be independent of $\beta$.
The continuum limit would be reached for 
$m_\ell, m_h \to 0$ with $m_\ell/m_h$ kept fixed. 
In particular, in the chiral limit ${m_\ell=0}$,
$m_h$ would only serve to set the scale.

There is some analogy between $m_h$ in our model 
and the bare coupling constant $g$ in QCD: near the 
fixed-point the physical values of masses and other 
observables do not depend on them.
In principle, in the chiral limit, 
the theory built in the neighborhood
of the IR fixed-point would be a self-consistent,
parameter free theory, very much like QCD with
massless quarks.

However, even if a theory built on the IR fixed-point were
self-consistent, like QCD it would have to be embedded into
a more general BSM framework.
Then, on various phenomenological grounds (scaling dimensions
of the condensate, of baryonic operators etc.), it can be argued
that the theory should be strongly coupled: thus,  still
within the conformal window, but as close as possible to its beginning. 

 There is some evidence that an SU(3) theory 
with 10 massless flavors may be conformal~\cite{Chiu:2016uui,Chiu:2017kza}. This makes it quite valuable
to perform an investigation, similar to the one presented
in these proceedings, but with 4 massless and 6 massive fermions.
Because of computational and, to some extent, 
theoretical difficulties~\cite{Hasenfratz:2017mdh},
however, a simulation with 10 flavors could not avail itself of the
simplifications brought about by staggered fermions.  Moreover
the need of preserving a clearly defined chiral limit practically
forces one to using the much more computationally demanding
domain-wall discretization.  This is currently planned by
the Lattice Strong Dynamics collaboration 
(http://lsd.physics.yale.edu/).  Hopefully, the efforts of
the LSD collaboration and/or some other group endowed with sufficient
computational resources will clarify, in a not too distant future,
the properties of the SU(3) four-plus-six fermion system.

\section*{Acknowledgments}
The authors thank their colleagues in the LSD Collaboration for fruitful and inspiring discussions. 
Computations for this work were carried out in part on facilities of the USQCD Collaboration, which are funded by the Office of Science of the U.S.~Department of Energy, on computers at the MGHPCC, in part funded by the National Science Foundation (award OCI-1229059), and on computers allocated under the NSF Xsede program to the project TG-PHY120002. 
We thank Boston University, Fermilab, the NSF and the U.S.~DOE for providing the facilities essential for the completion of this work.  A.H. acknowledges support by DOE grant 
DE-SC0010005 and C.R. by DOE grant DE-SC0015845.   This project has received funding from the European Union's Horizon 2020 research and innovation programme under the Marie Sk{\l}odowska-Curie grant agreement No 659322.

\bibliography{BSM}

\begin{thebibliography}{21}

\bibitem{Luty:2004ye}
M.A. Luty, T.~Okui, JHEP \textbf{09}, 070 (2006), \texttt{hep-ph/0409274}

\bibitem{Dietrich:2006cm}
D.D. Dietrich, F.~Sannino, Phys. Rev. \textbf{D75}, 085018 (2007),
  \texttt{hep-ph/0611341}

\bibitem{Vecchi:2015fma}
L.~Vecchi (2015), \texttt{1506.00623}

\bibitem{Ferretti:2013kya}
G.~Ferretti, D.~Karateev, JHEP \textbf{03}, 077 (2014), \texttt{1312.5330}

\bibitem{Ma:2015gra}
T.~Ma, G.~Cacciapaglia, JHEP \textbf{03}, 211 (2016), \texttt{1508.07014}

\bibitem{Brower:2014dfa}
R.C. Brower, A.~Hasenfratz, C.~Rebbi, E.~Weinberg, O.~Witzel, J. Exp. Theor.
  Phys. \textbf{120}, 423 (2015), \texttt{1410.4091}

\bibitem{Brower:2015owo}
R.C. Brower, A.~Hasenfratz, C.~Rebbi, E.~Weinberg, O.~Witzel, Phys. Rev.
  \textbf{D93}, 075028 (2016), \texttt{1512.02576}

\bibitem{Hasenfratz:2016gut}
A.~Hasenfratz, C.~Rebbi, O.~Witzel, Phys. Lett. \textbf{B773}, 86 (2017),
  \texttt{1609.01401}

\bibitem{Hasenfratz:2016uar}
A.~Hasenfratz, C.~Rebbi, O.~Witzel, PoS \textbf{LATTICE2016}, 226 (2016),
  \texttt{1611.07427}

\bibitem{Cheng:2013bca}
A.~Cheng, A.~Hasenfratz, G.~Petropoulos, D.~Schaich, PoS \textbf{LATTICE2013},
  088 (2013), \texttt{1311.1287}

\bibitem{Cheng:2013xha}
A.~Cheng, A.~Hasenfratz, Y.~Liu, G.~Petropoulos, D.~Schaich, Phys.Rev.
  \textbf{D90}, 014509 (2014), \texttt{1401.0195}

\bibitem{Hasenfratz:2001hp}
A.~Hasenfratz, F.~Knechtli, Phys. Rev. \textbf{D64}, 034504 (2001),
  \texttt{hep-lat/0103029}

\bibitem{Hasenfratz:2007rf}
A.~Hasenfratz, R.~Hoffmann, S.~Schaefer, JHEP \textbf{0705}, 029 (2007),
  \texttt{hep-lat/0702028}

\bibitem{Agashe:2014kda}
K.A. Olive et~al. (Particle Data Group), Chin. Phys. \textbf{C38}, 090001
  (2014)

\bibitem{Aoki:2012eq}
Y.~Aoki, T.~Aoyama, M.~Kurachi, T.~Maskawa, K.i. Nagai, H.~Ohki, A.~Shibata,
  K.~Yamawaki, T.~Yamazaki ({LatKMI}), Phys. Rev. \textbf{D86}, 054506 (2012),
  \texttt{1207.3060}

\bibitem{Fodor:2011tu}
Z.~Fodor, K.~Holland, J.~Kuti, D.~Nogradi, C.~Schroeder, K.~Holland, J.~Kuti,
  D.~Nogradi, C.~Schroeder, Phys. Lett. \textbf{B703}, 348 (2011),
  \texttt{1104.3124}

\bibitem{Aoki:2013zsa}
Y.~Aoki, T.~Aoyama, M.~Kurachi, T.~Maskawa, K.i. Nagai, H.~Ohki, E.~Rinaldi,
  A.~Shibata, K.~Yamawaki, T.~Yamazaki (LatKMI), Phys. Rev. Lett. \textbf{111},
  162001 (2013), \texttt{1305.6006}

\bibitem{Dowdall:2013rya}
R.J. Dowdall, C.T.H. Davies, G.P. Lepage, C.~McNeile, Phys. Rev. \textbf{D88},
  074504 (2013), \texttt{1303.1670}

\bibitem{Chiu:2016uui}
T.W. Chiu (2016), \texttt{1603.08854}

\bibitem{Chiu:2017kza}
T.W. Chiu, PoS \textbf{LATTICE2016}, 228 (2017)

\bibitem{Hasenfratz:2017mdh}
A.~Hasenfratz, C.~Rebbi, O.~Witzel, \emph{{Testing Fermion Universality at a
  Conformal Fixed Point}} (2017), \texttt{1708.03385},
  \urlstyle{tt}\url{http://inspirehep.net/record/1615760/files/arXiv:1708.03385.pdf}

\end{thebibliography}

\end{document}